\documentclass[aps,prb,preprint,showpacs,amsmath,floatfix,superscriptaddress]{revtex4}

\usepackage{graphicx}
\usepackage{dcolumn}
\usepackage{bm}
\usepackage{accents}
\usepackage{xcolor}
\usepackage{epstopdf}
\begin{document}

\title{A Locally-Preferred Structure Characterizes All Dynamical Regimes of a Supercooled Liquid}

\author{Ryan Soklaski}
\affiliation{Department of Physics and Institute of Materials
Science and Engineering, Washington University in St. Louis, St.
Louis, MO 63130, USA}
\author{Vy Tran}
\affiliation{Department of Physics and Institute of Materials
Science and Engineering, Washington University in St. Louis, St.
Louis, MO 63130, USA}
\author{Zohar Nussinov}
\affiliation{Department of Physics and Institute of Materials
Science and Engineering, Washington University in St. Louis, St.
Louis, MO 63130, USA}
\affiliation{Department of Condensed Matter
Physics, Weizmann Institute of Science, Rehovot, 76100 Israel}
\author{K.F. Kelton}
\affiliation{Department of Physics and Institute of Materials
Science and Engineering, Washington University in St. Louis, St.
Louis, MO 63130, USA}
\author{Li Yang}
\affiliation{Department of Physics and Institute of Materials
Science and Engineering, Washington University in St. Louis, St.
Louis, MO 63130, USA}
\date{}

\begin{abstract}
Recent experimental results suggest that metallic liquids
universally exhibit a high-temperature dynamical crossover, which
is correlated with the glass transition temperature ($T_{g}$). We
demonstrate, using molecular dynamics results for
$\mathrm{Cu_{64}Zr_{36}}$, that this temperature, $T_{A} \approx 2
\times T_{g}$, is linked with cooperative atomic rearrangements
that produce domains of connected icosahedra. Supercooling to a
new characteristic temperature, $T_{D}$, is shown to produce
higher-order cooperative rearrangements amongst connected
icosahedra, which manifests as the formation of large Zr-rich connected domains
that possess macroscopic proportions of the liquid's icosahedra.
This coincides with the decoupling of atomic diffusivities,
large-scale domain fluctuations, and the onset of glassy dynamics in the liquid.
These extensive domains then abruptly
stabilize above $T_{g}$ and eventually percolate before the glass
is formed. All characteristic temperatures ($T_{A}$, $T_{D}$ and
$T_{g}$) are thus connected by successive manifestations of the
structural cooperativity that begins at $T_{A}$.
\end{abstract}

\maketitle

\section{Introduction}

As a liquid is supercooled below its melting temperature, the
characteristic timescales of its dynamics become increasingly
stretched. In a dramatic departure from exhibiting simple
diffusive particle dynamics, its viscosity, $\eta$, increases
rapidly as it develops general, glassy dynamic features
\cite{1970Johari, 2000Angell, 2009Cavagna, 2009Gotze}. Efforts to
universally describe these phenomena as a liquid is supercooled to
its glass transition temperature ($T_{g}$) remain divided in both
their approaches and successes \cite{Karmakar2014, 2009Cavagna,
2009Gotze, Kirkpatrick1989, Debendetti2001, Kivelson2008, Das2004,
Angell2008, Tarjus2005,Tanaka2000, Aharanov2007}. In these
theories, the roles played by structure, if any, are a major point
of discrepancy. The molecular dynamics (MD) results presented here
reveal a striking correlation between structure and dynamics
across a broad temperature range. In particular, a high
temperature structural crossover is shown to underlay a recently
reported dynamical one \cite{Kelton2014}.

Numerous works point to the fundamental importance of
locally-preferred structures (LPS)s in glass forming liquids
\cite{Tarjus2005, Tanaka2010, Chen1988, Baumer2013, Malins2013}. One such
thermodynamic theory describes the growth of frustration-limited
domains (FLD)s of the LPS as the liquid is cooled
\cite{Tarjus2005, Kivelson1995, Nussinov2004}. This FLD theory accurately
describes glassy relaxation processes and predicts a temperature,
$T_{A}$, below which a fragile liquid develops a super-Arrhenius
temperature dependence for $\eta$. Very recently, an analysis of
experimental data for a wide variety of metallic liquids revealed
that $T_{A}$ correlates strongly with $T_{g}$, such that $T_{A}
\approx 2 \times T_{g}$ \cite{Kelton2014}. This temperature may
correspond to a crossover phenomenon created by competition
between configurational excitations and phonons \cite{Iwashita2013}.
Remarkably, it was found that, by using $T_{A}$ and
particle density as scaling factors, the viscosity data for all
strong and fragile metallic liquids fall on a universal curve
across a broad temperature range \cite{Kivelson1995,Kelton2014}.

Despite the merits of the FLD theories, an important problem
persists.  The measured structural changes of the liquid across
the accessible supercooled temperature range are extremely subtle,
casting doubts on the extent and even the existence of the FLDs
\cite{2009Cavagna, Kivelson2008} as well as the structural changes
that accompany processes at $T_{A}$ \cite{Chen1988, Iwashita2013}.
Therefore, observing FLDs and studying their role in the dynamical
crossover at $T_{A}$ would bolster FLD theory, the interpretation
of the apparent universality across metallic liquids, and the
overall understanding of the relationship between structure and
dynamics in liquids

In this work, we demonstrate the development and growth of FLDs in
a metallic liquid as it is cooled through $T_{A}$, providing a
coherent picture of the structural and dynamical features of the
cooperativity that arises at this crossover temperature as well as
within the supercooled regime. Rapidly growing fluctuations in
domain sizes are linked to the development of glassy dynamics,
leading to the identification of another crossover at lower
temperature, $T_{D}$, which may be possible to observe in
experimental structural studies \cite{Mauro2013, Mauro2014}.
While supercooling, these extensive domain then quickly stabilize,
marked by a sudden drop in domain fluctuations, and percolate across
the system before the temperature reaches $T_{g}$. The MD results presented here suggest that all of
these characteristic temperatures, $T_{A}$, $T_{D}$ and $T_{g}$
correspond to a cascade of cooperative structural rearrangements
involving a LPS that begins at $T_{A}$, a result consistent with
the prediction from recent experimental studies of viscosity
\cite{Kelton2014}.

The remainder of this paper is organized as follows: in section
II, we introduce the methods of our MD simulations. In
section III, we discuss our methods of our Voronoi analysis, and introduce
definitions of icosahedral ordering. Section IV introduces the temperatures
$T_{A}$ and $T_{D}$, and the dynamical features that accompany them.
Section V defines liquid relaxation and restructuring timescales and discusses the role
played by cooperative structural rearrangements at $T_{A}$ and $T_{D}$.
Here also we show that $T_{A}$ marks the onset of connected icosahedral ordering.
Section VI links that dynamical features that begin at $T_{D}$ with the rapid
proliferation of connected domains of icosahedra. Section VII provides an overview
of the structure-dynamic relationships of the liquid and its icosahedron domains from high temperatures
down to $T_{g}$, where were find that a single domain of icosahedra has percolated the system.
Section VIII contains a summary of our findings.

\section{Methods}
Results were obtained from classical MD simulations
of the metallic liquid $\mathrm{Cu_{64}Zr_{36}}$, which among best glass forming compositions
of this alloy \cite{Bendert2012}.
All MD simulations were conducted using LAMMPS \cite{lammps} with
GPU-parallelization packages \cite{gpu}. Atomic interactions
in the Cu$_{64}$Zr$_{36}$ system, including approximations of many-body
interactions, are described using the Finnis-Sinclair generalization of the
embedded atom method \cite{EAM}. This semi empirical potential was selected for its
ability to accurately reproduce structures and properties of Cu$_{64}$Zr$_{36}$ in
both the liquid and glass regimes, as confirmed by comparisons made with x-ray diffraction data and ab initio calculation results \cite{Mendelev2009}. The reported
calorimetric glass formation temperature ($T_{g}$) for this potential is
approximately $750 K$. We confirm this by measuring the system
volume, given zero average pressure, as a function of temperature,
and identifying the temperature at which the thermal expansion coefficient
changes.

Isothermal, isobaric simulations (sampled from ensembles with
constant $N$, $\langle P \rangle$, and $\langle T \rangle$
\cite{NPT}) were conducted with $\langle P \rangle = 0$ and
$N=3\times 10^{4}$ atoms, using periodic boundary conditions, the
Verlet integration method, and the Nose-Hoover thermostatting
method. Each simulation was initialized with a random atomic
configuration that evolves at $T= 3300 K$ for $0.5 ns$ to achieve
equilibrium. The liquid was then quenched at a rate of $10^{11}
K/s$ down to its target temperature and subsequently relaxed for
$20 ns$ before structure and dynamics data were collected. The
integration timestep during the quenching and relaxation processes
was $5 fs$. The integration step was then decreased to $2 fs$, and
data was collected for each timestep. For a given target
temperature, atomic-level data, such as the positions and
velocities were recorded across a $0.2 ns$ interval; the system's
Cauchy stress tensor was recorded for $8 ns$. Additional
independent simulations were conducted for all temperatures below
$1100 K$ in addition to $2200 K$ and $2500 K$. To check for system
size effects, additional simulations of $N=1\times 10^{4}$ atoms
were performed.

\section{Icosahedral Ordering and Frustration-Limited Domains}

Voronoi analysis utilizing a weighted-bisector method was
performed utilizing the Voro++ software library
\cite{Rycroft2006}. Small faces comprising less than $0.5\%$ of a
Voronoi cell's surface area were removed. The analysis was
performed on each time step for which atomic positions were
recorded.
 The LPS for Cu$_{64}$Zr$_{36}$ is an icosahedron-shaped
cluster of atoms \cite{Liu2013, Cheng2008, Ding2014, Sheng2006},
consisting of a Cu atom with 12 nearest neighbor atoms, which form
a weighted-Voronoi cell with 12 pentagonal faces.
Distorted icosahedra, which contain some
non-pentagonal Voronoi faces, are treated as distinct structures
and are not considered to be LPSs. FLDs are thus comprised of
connected icosahedra. Two icosahedra are connected if they
share a vertex, an edge, a face, or they interpenetrate, meaning
that they share 1, 2, 3, or 7 atoms, respectively
\cite{Soklaski2013}.

\begin{figure}[t]
\includegraphics[scale=0.4]{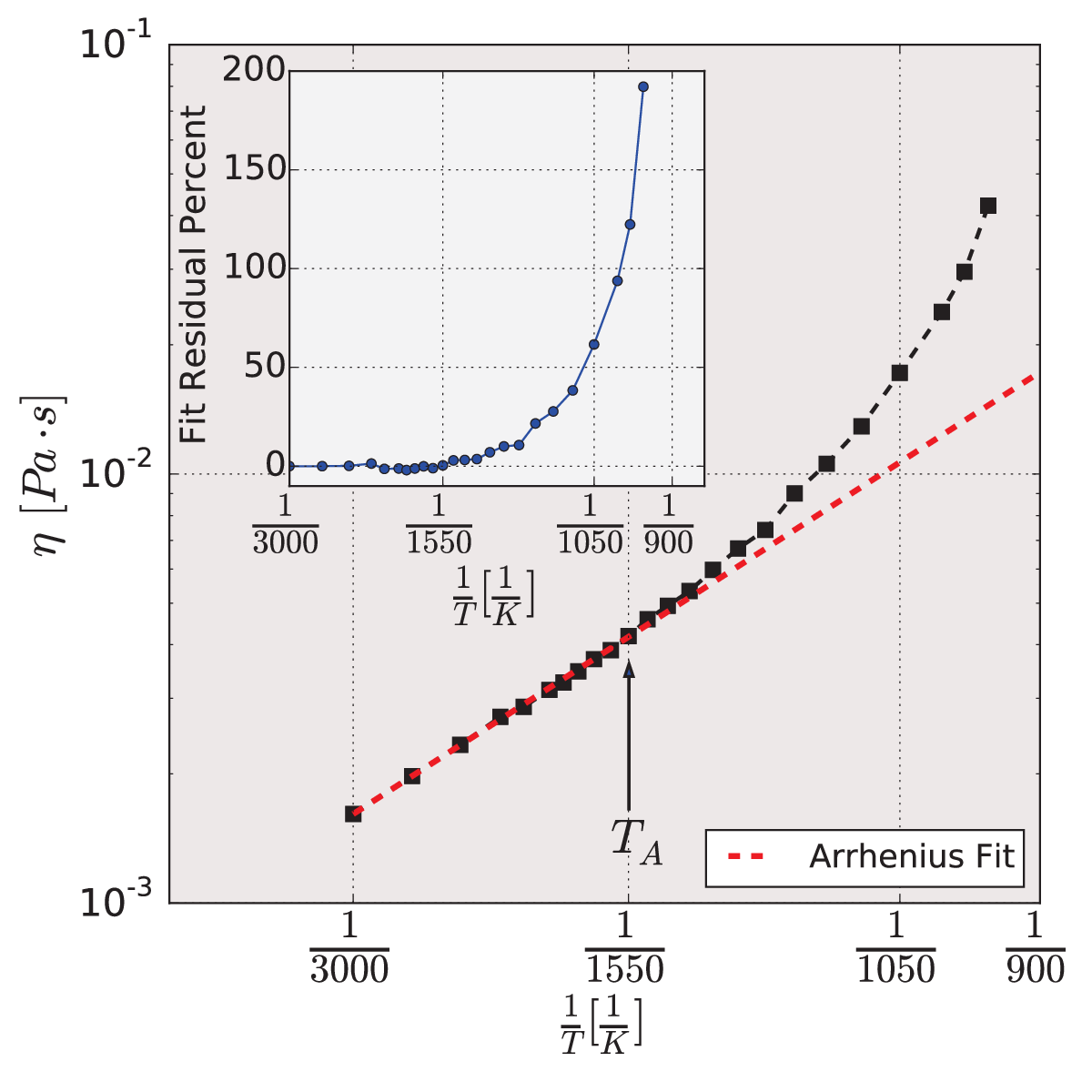}
\caption{(Color online) Viscosity, $\eta$, as a function of
$\frac{1}{T}$. $\eta$ begins to exhibit super-Arrhenius growth
below $T_{A} = 1550 K$.  (Inset) The percent residual of $\eta$
from the Arrhenius fit.} \label{fig:viscosity}
\end{figure}

\section{Dynamical features of liquid $Cu_{64}Zr_{36}$: Break down of Stokes-Einstein and Decoupling of Diffusivities}

Before analyzing the structural evolution of
$\mathrm{Cu_{64}Zr_{36}}$, a summary of the dynamical features
that develop in the liquid across a broad temperature range is presented.
Viscosity ($\eta$) data obtained using the Green-Kubo relationship for the
stress tensor autocorrelation function \cite{Hansen2006}
are shown in the main panel of Figure~\ref{fig:viscosity}. A
change from an Arrhenius temperature dependence to a
super-Arrhenius form occurs when the liquid is cooled below $1550
K$, which is therefore identified as $T_{A}$. In accord with
recent experimental study \cite{Kelton2014}, this value is near $2
\times T_{g}$ ($T_{g} \approx 750 K$ \cite{Mendelev2009}).

\begin{figure}[t]
\includegraphics[scale=0.43]{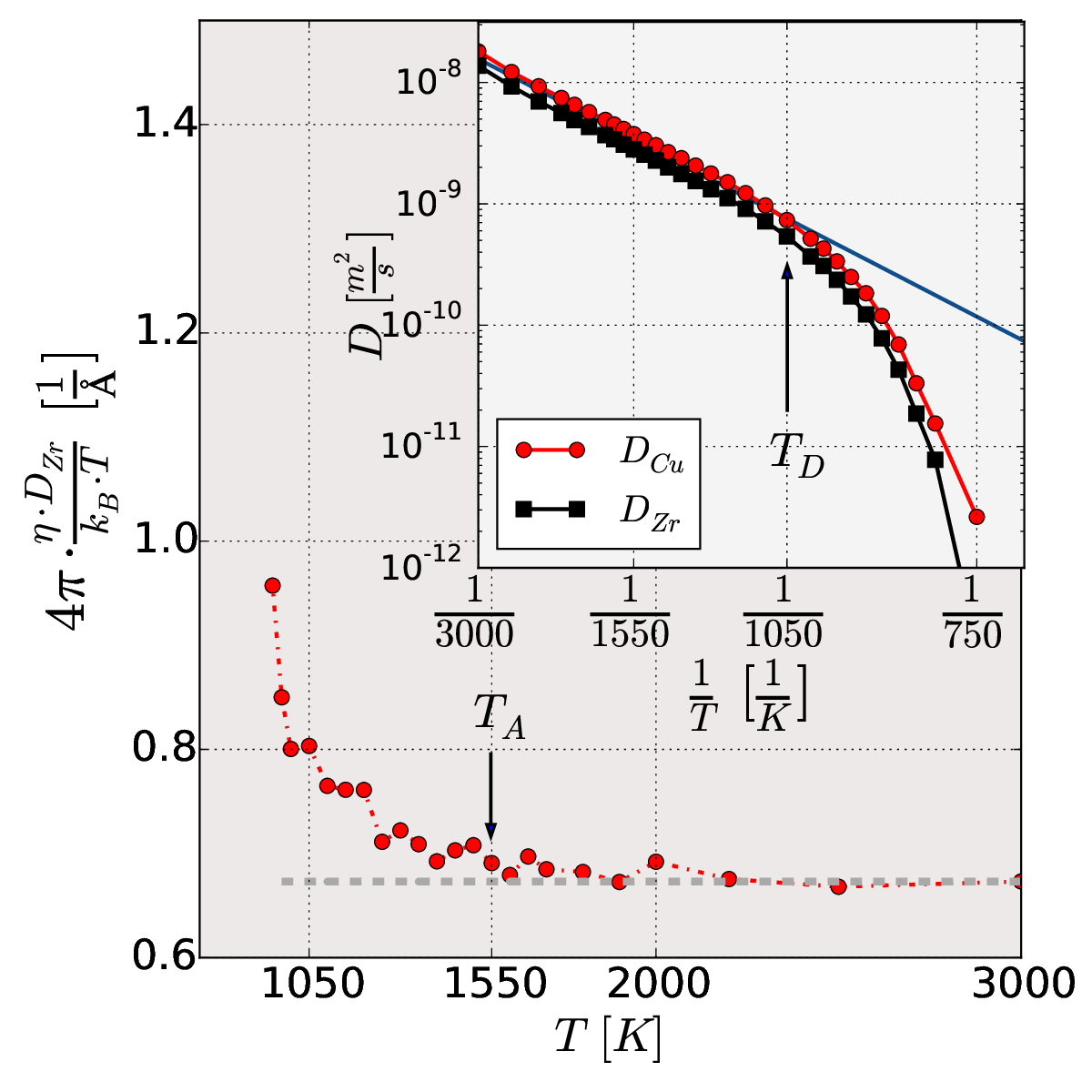}
\caption{(Color online) The Stokes-Einstein ratio as a function of
$T$. The Stokes-Einstein relationship becomes violated below $T_{A}$
(Inset) Cu and Zr diffusivities versus $\frac{1}{T}$. A
transition from a high-\textit{T} linear fit on a log scale (solid line)
occurs at $T_{D} = 1050 K$ for both species.}
\label{fig:diffusivity}
\end{figure}

The liquid diffusivity data, plotted in the inset of
Figure~\ref{fig:diffusivity}, maintains an Arrhenius form through
lower temperatures than does $\eta$, indicating that $T_{A}$ also
marks the breakdown of the Stokes-Einstein relationship for Zr
atoms,
\begin{equation}\label{eq:stokes} \eta \cdot D_{Zr} \propto \frac{k_{B} \cdot T}{R_{Zr}}, \end{equation}
which are the larger diffusing solute species in the system
\cite{2009Cavagna,Inoue2008}. Here, $D_{Zr}$ is the diffusion
coefficient of tagged Zr particles as measured from the long-time
integral of their velocity autocorrelation functions, $T$ is the
temperature, and $R_{Zr}$ is the effective particle radius. The
data presented in the main panel of Figure~\ref{fig:diffusivity}
shows that for high temperatures the Stokes-Einstein ratio is
roughly a constant 0.67, which is comparable to $\frac{1}{R_{Zr}}
\approx 0.65\ 1/\AA$ \cite{radius}. At $T_{A}$, the accelerated
growth in $\eta$ causes a deviation from this constant, indicating
that the liquid is departing from its ``simple" behavior, and is
developing dynamical heterogeneities \cite{Karmakar2014,
Cicerone1996}. Specifically, slow-moving regions develop in the
liquid that dictate the timescale of structural relaxations, as
measured by $\eta$, whereas fast-moving regions control the
dynamical timescale proportional to $\frac{1}{D}$.

\begin{figure}[t]
\includegraphics[scale=0.43]{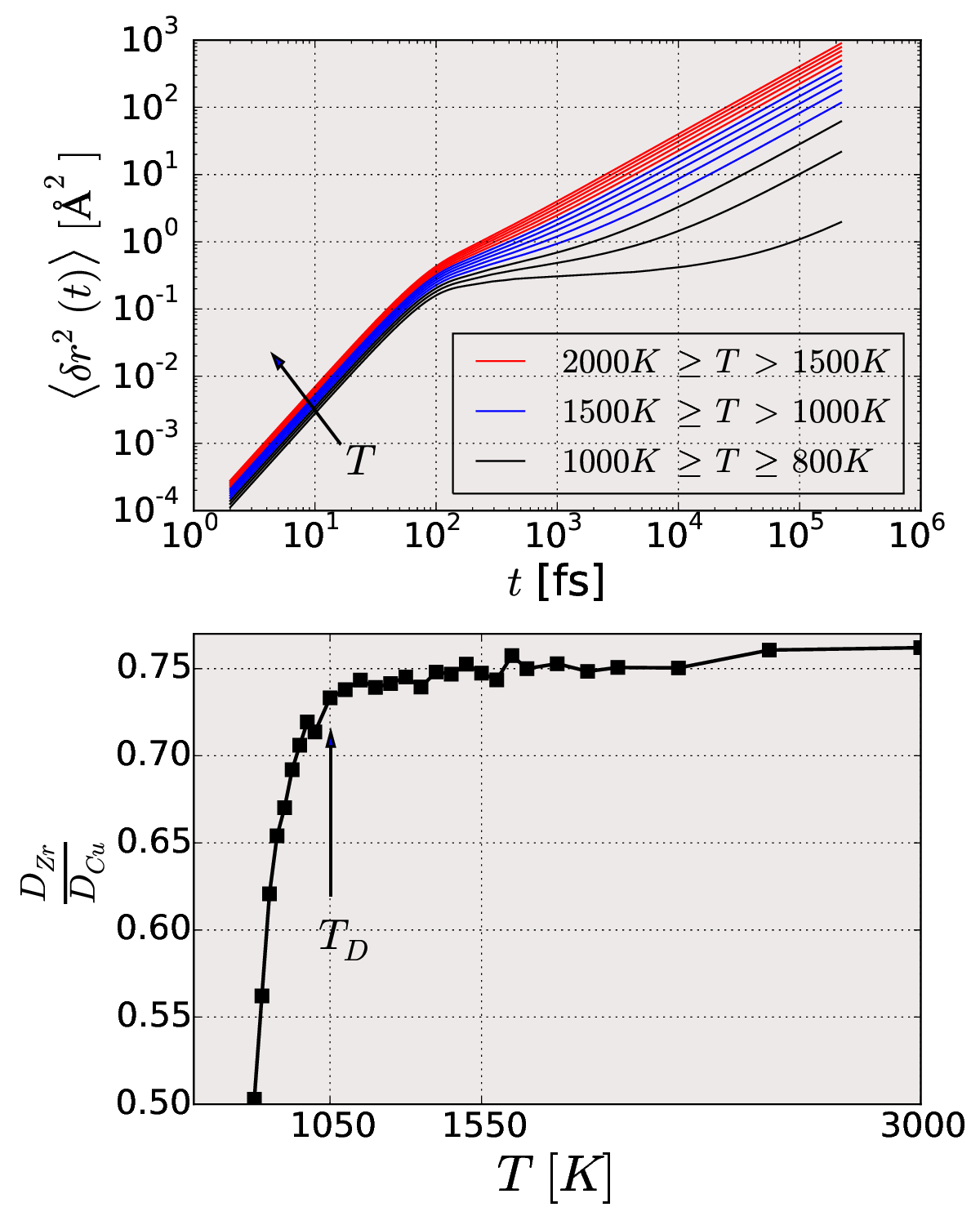}
\caption{(Color online) (Top) The mean-squared displacement
time-trajectories of atoms for various liquid temperatures on a
log-log scale. The temperatures included are separated by $100K$
intervals. A distinct plateau feature, characteristic of glassy
dynamics, develops for $T \lesssim T_{D}$. (Bottom) The ratio of
the diffusivity of Zr atoms to that of Cu atoms. The diffusivities
are found to decouple sharply for $T \leq T_{D}$.}
\label{fig:MSD_DRatio}
\end{figure}

It is striking to find that the Stokes-Einstein relationship is
violated at such a high temperature - literature reviews of
supercooled liquids broadly indicate that this relationship is
expected to hold well into a liquid's supercooled regime
\cite{2009Cavagna,Debendetti2001}. However, this prescription is
chiefly informed by empirical studies of molecular liquids, such
as \textit{o}-terphenyl, where the violation occurs near $1.2
\times T_{g}$ (near the mode-couple temperature)
\cite{Mapes2006,Swallen2003}. Our finding, that the
Stokes-Einstein relationship is violated well above $1.2 \times
T_{g}$, are supported by several empirical studies of metallic
liquids \cite{Brillo2011,Brillo2008,Meyer2003}. Furthermore, an
illuminating concurrent MD study of liquid
Cu$_{40}$Zr$_{51}$Al$_{9}$ submitted very shortly after the
original submission of the current work similarly found that the
relation breaks down specifically at $T_{A} \approx 2 \times
T_{g}$ for that alloy as well \cite{egami2015}. Taken together,
these results bolster our finding of Stokes-Einstein violation at
$T_{A}$ to a broader range of materials

It is important to note that the local violation of the
Stokes-Einstein equation at $T_{A}$ does not signify the onset of
strong glassy dynamics in $\mathrm{Cu_{64}Zr_{36}}$. That is, the
system does not yet exhibit plateau-separated fast ($\beta$) and
slow ($\alpha$) relaxation processes at $T_{A}$. This can be seen
in the inset of Figure~\ref{fig:diffusivity}, which displays the
temperature dependence of the diffusivities for Cu and Zr
particles, and in the top panel of Figure~\ref{fig:MSD_DRatio},
which contains a series of mean-squared displacement (MSD) time
trajectories for several temperatures. The MSD trajectory at
$T_{A}$ exhibits an immediate transition between simple ballistic
and diffusive motion regimes without any plateau feature. This
changes once the diffusivity, $D(T)$, departs from an
exponentially-decaying function of $1/T$ at $1050 K$, which we
label $T_{D}$. $D$ reflects the long-time asymptotic behaviors of
the particles' MSDs: $\langle\delta r(t)^{2}\rangle \sim 6Dt\ (t
\rightarrow \infty)$, thus the accelerated decline in $D(T)$ for
both atomic species reflects the development of a sustained
plateau that separates two-step relaxation processes in the
supercooled liquid (Fig~\ref{fig:MSD_DRatio}), as well as growing
diffusive-motion activation energy barriers. This is
characteristic of caged particle dynamics and is an important
equilibrium signature of the impending glass transition
\cite{2009Gotze, 2009Cavagna}.

Also at $T_{D}$, the motions of Cu and Zr atoms begin to decouple,
leading to a sharp decrease in $\frac{D_{Zr}}{D_{Cu}}$ with
decreasing temperature; this is shown in the lower panel of
Figure~\ref{fig:MSD_DRatio}. Here, the system's energy landscape
begins to to disproportionately impede the activation of the
larger Zr atoms, whereas above $T_{D}$, the two species diffuse
via the same physical mechanism with $\frac{D_{Zr}}{D_{Cu}}
\approx \frac{R_{Cu}}{R_{Zr}}$ \cite{radius}. This decoupling
strongly suggests that a structural feature begins to emerge at
$T_{D}$ that skews the supercooled liquid's dynamics. We will
investigate this underlying structure in section VI. Lastly, we
note that a recent empirical study of a Zr-based metallic liquid
reports a similar decoupling of dynamics at a temperature
comparable to $T_{D}$ \cite{Basuki2014}.

In summary, $T_{A}$ ($1550K$) is the temperature below which
$\eta$ begins to exhibit super-Arrhenius growth, and where the
Stokes-Einstein relationship first breaks down in the liquid.
$T_{D}$ ($1050K$) then marks the early onset of stretched, glassy
dynamical features in the supercooled liquid, which appear to be
shaped by an emerging robust structural feature.

\section{$T_{A}$ and Cooperative structural rearrangements}

\begin{figure}[t]
\includegraphics[scale=0.38]{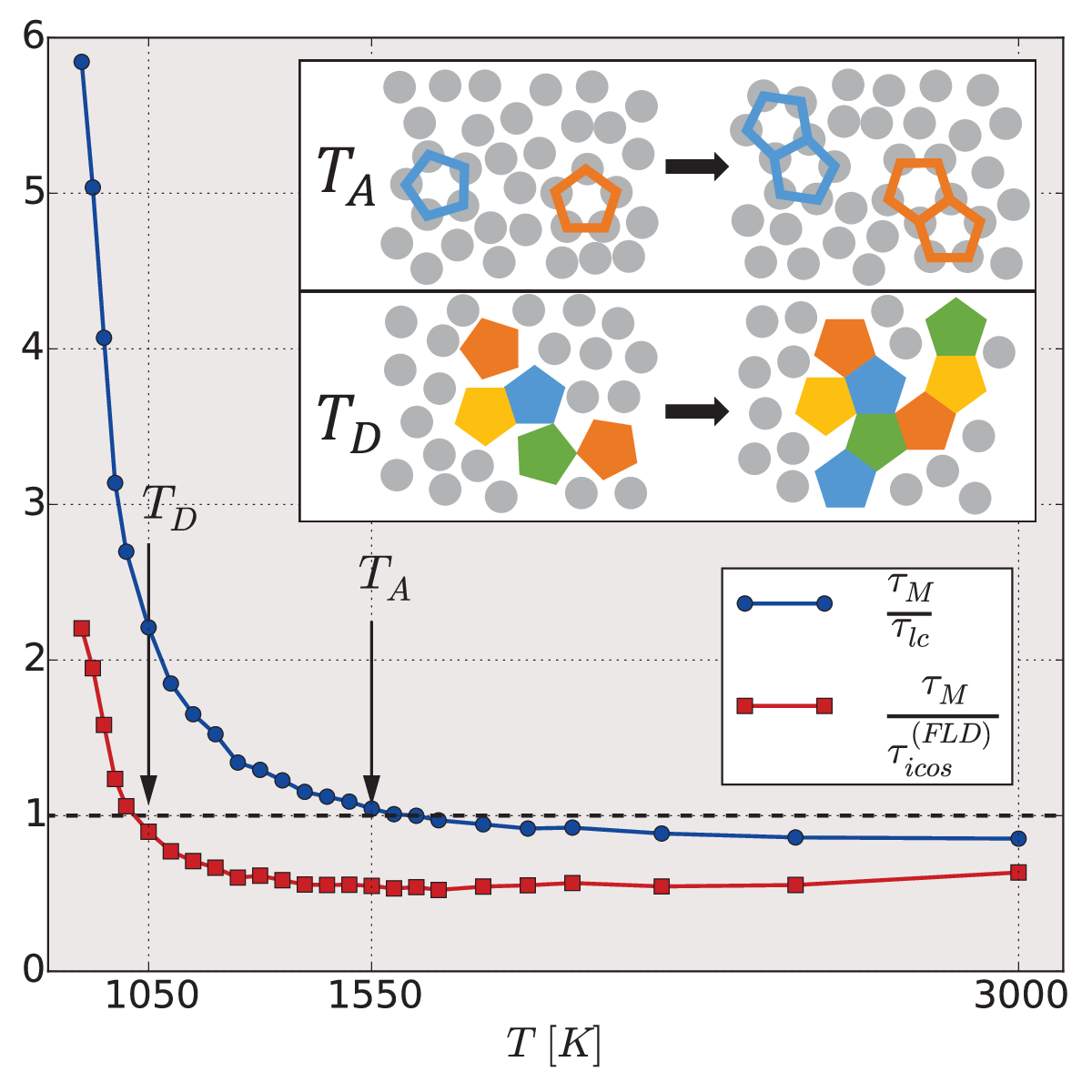}
\caption{(Color online) Ratio of the Maxwell time to the local
cluster time and the domain-connected icosahedron lifetime,
respectively. $\tau_{M}$ surpasses $\tau_{LC}$ and
$\tau^{(FLD)}_{icos}$ at $T_{A}$ and $T_{D}$, respectively.
(Inset) A 2D illustration of the cooperative rearrangements that
begin contributing to the liquid relaxation process at $T_{A}$ and
at $T_{D}$, respectively. Here the 2D pentagons represent 3D
icosahedral clusters of atoms.} \label{fig:timescales}
\end{figure}

The crossover from the ``simple liquid" regime at $T_{A}$ marked
by the development of super-Arrhenius relaxation times, and the
development of glassy dynamics at $T_{D}$ both coincide with the
onset of structural cooperativity associated with FLDs of
connected icosahedra. Our analysis of this cooperativity follows
and expands on the work of Iwashita, Nicholson, and Egami
\cite{Iwashita2013} and introduces three pertinent timescales.
First, the Maxwell time, or shear stress relaxation time,
\begin{equation}\label{eq:maxwell} \tau_{M} \equiv \frac{ \int_{0}^{\infty} \langle\sigma_{ij}(t)\sigma_{ij}(0)\rangle dt }{\langle\sigma_{ij}(0)^2\rangle}, \end{equation}
provides a timescale during which the liquid behaves like a solid
and exhibits an elastic shear response without flowing
\cite{2009Cavagna,Inoue2008}. $\sigma_{ij}(t)$ is the
time-dependent Cauchy stress tensor \cite{Inoue2008}. Second, the
local cluster time, $\tau_{LC}$, is defined as the average time
required for an atom to change its coordination number, i.e. the
time needed for an atom to lose or gain a Voronoi neighbor. In
addition to these two times, we introduce the icosahedron
lifetime, $\tau_{icos}$, which is the average time required for an
icosahedron to lose a vertex atom, gain an extraneous vertex atom,
or distort its shape. $\tau_{icos}$ for an icosahedron involved in
a domain via interpenetrating connection is labelled
$\tau^{(FLD)}_{icos}$. Measurements of $\tau_{LC}$ and
$\tau_{icos}$ provide a context for understanding the timescales
over which atomic rearrangements of the liquid structure occur.

\begin{figure}[t]
\includegraphics[scale=0.38]{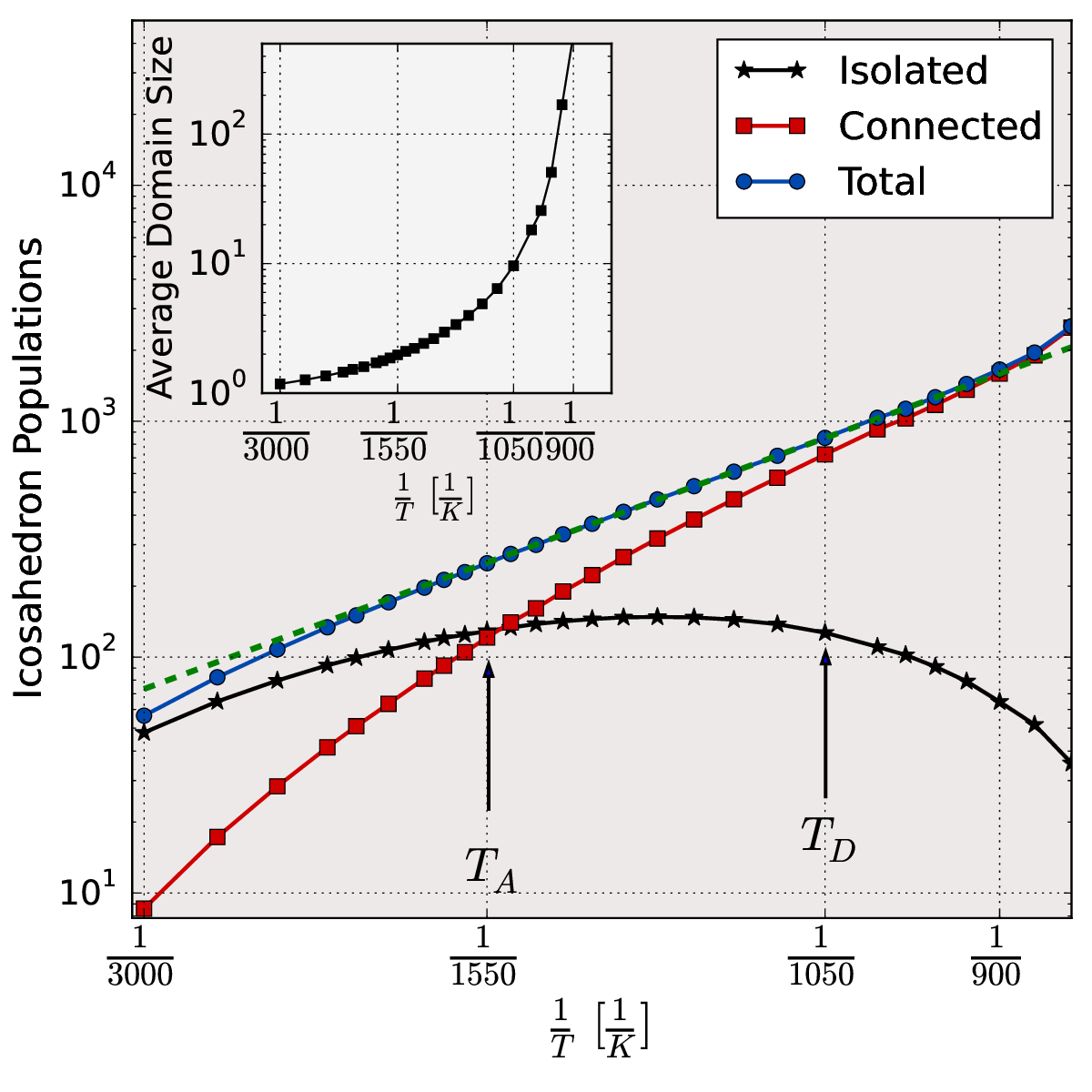}
\caption{(Color online) The populations of connected and isolated
icosahedra, plus their sum. A low-$T$ linear fit on the log scale
is indicated with a dashed line. A crossover in the isolated and
connected populations occurs as the liquid is cooled throug $T_{A}$.
(Inset) The average
connected-icosahedron domain size, $\bar s$, on a log scale.
A domain's size is given by the number of icosahedra participating
in a singly-connected structure} \label{fig:icos_pop}
\end{figure}

The primary portion of Figure~\ref{fig:timescales} shows the
ratios of $\tau_{M}$ to $\tau_{LC}$ and $\tau^{(FLD)}_{icos}$,
respectively, as a function of temperature. Liquid relaxation in
the high temperature, simple liquid regime $(T>T_{A})$ is
characterized by uncorrelated local rearrangements of atomic
clusters: $\tau_{M} \approx \tau_{LC}$. That the high temperature
liquid relaxation time is generally controlled by $\tau_{LC}$ in
atomic liquids was first shown by Iwashita \textit{et al.}
\cite{Iwashita2013}. Here, $\tau_{M}$ is so short that consecutive
atomic rearrangements occur only after the surrounding environment
has relaxed in response to shear strains. It is not until the
system is cooled through $T_{A}$ that $\tau_{M} > \tau_{LC}$, and
that typical atomic clusters can begin to rearrange with one
another on a time scale during which the liquid behaves like a
solid. Here cooperative restructuring can begin contributing to
the liquid's relaxation process \cite{Iwashita2013}.

What is produced by these cooperative rearrangments? That is, what
difference is there between the average structure of the liquid at
temperatures above $T_{A}$ and those below? Measuring the the
populations of the isolated, and connected icosahedra in a volume
containing $3\times 10^{4}$ atoms reveals that a structural
crossover occurs at $T_{A}$: above $T_{A}$ it is more likely to
find isolated icosahedra than connected ones, and below $T_{A}$
connected icosahedra become the dominant population. This can be
seen in Figure~\ref{fig:icos_pop}, which shows, on a log scale,
the evolution of these icosahedron populations as well as the
total number of icosahedra. The total population of icosahedra
proceeds to grow exponentially with inverse temperature for $T_{A}
> T$. In the context of our timescales, we find that in the regime
where $T_{A} > T > T_{D}$ icosahedra are stable in comparison to
the average local cluster configuration ($\tau^{FLD}_{icos} >
\tau_{LC}$), and are relatively robust on the timescale of liquid
relaxation ($\tau^{FLD}_{icos} \gtrsim \tau_{M}$) across this
temperature range. Thus the liquid relaxation process chiefly
involve non-icosahedral atomic clusters, which cooperatively
rearrange amidst the relatively inert icosahedra and form off of
them, new, connected icosahedra. This process is depicted in a
simplified 2D illustration in the inset of
Figure~\ref{fig:timescales}.

It is quite remarkable that the distinct dynamical features
associated with $T_{A}$ are linked with the incipient icosahedral
ordering in this system. While the structure-dynamics features of
icosahedron networks are becoming well-documented for temperatures
near $T_{g}$ \cite{Cheng2008,Ding2014,Mendelev2015}, the roles
played by these structures at higher temperatures have remained
largely unknown. Thus our findings stand amongst the first
explicit accounts of the structural features that accompany the
emergence of solidlike features in high temperature metallics
liquids. They also correspond closely with the narrative of FLD
theory, which predicts the development of frustrated domains at
$T_{A}$ in association with an avoided critical point in the
system \cite{Tarjus2005, Kivelson1995, Nussinov2004}. We proceed
by further tracking this icosahedral ordering into the supercooled
region, where prominent structural features are expected to emerge
near $T_{D}$.

\section{$T_{D}$ and the rapid onset of dominant icosahedral ordering}

\begin{figure}[t]
\includegraphics[scale=0.38]{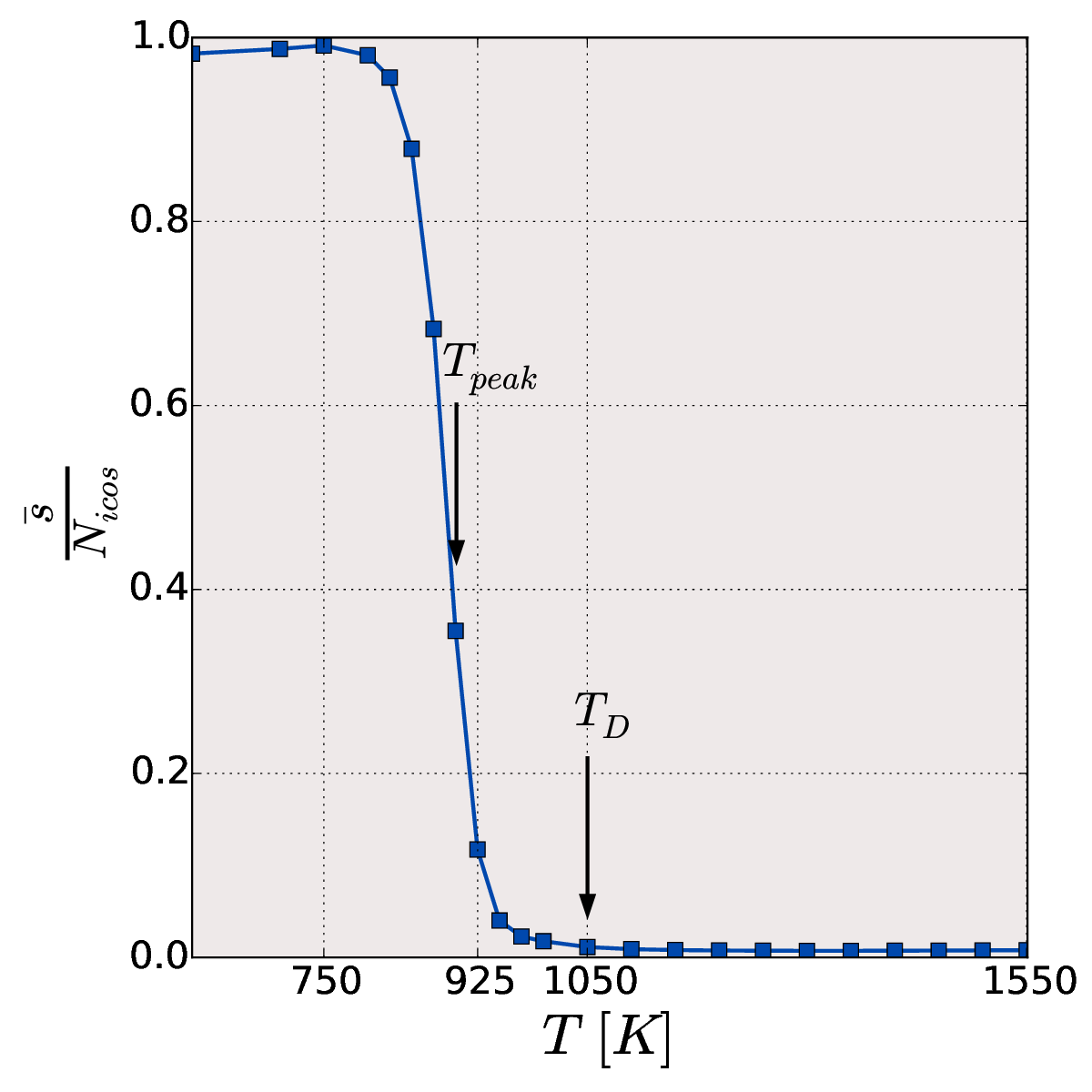}
\caption{(Color online) The fraction of the total number
of icosahedra in a 30,000 atom system, $N_{icos}$, contained
in an FLD. Below $T_{D}$, a domain
contains a macroscopic number of the system's icosahedra.
$T_{peak}$ marks the maximum susceptibility of the fractional
occupation.} \label{fig:perc_net}
\end{figure}

As the liquid is supercooled through $T_{D}$, its relaxation time
grows such that $\tau_{M} \gg \tau^{(FLD)}_{icos}$  for $T <
T_{D}$ (see Fig~\ref{fig:timescales}), and thus its relaxation
processes begin to involve consecutive structural rearrangements
of connected icosahedra within FLDs. The illustration in
Figure~\ref{fig:timescales} depicts a simple schematic of such
processes. We find that the onset of this medium-ranged
cooperative restructuring accompanies the enhancement of the
diffusive-motion energy barrier and the early development of
general glassy dynamics observed at $T_{D}$, and it leads to a
dramatic proliferation of highly connected FLDs of icosahedra
throughout the liquid.

\begin{figure}[t]
\includegraphics[scale=0.38]{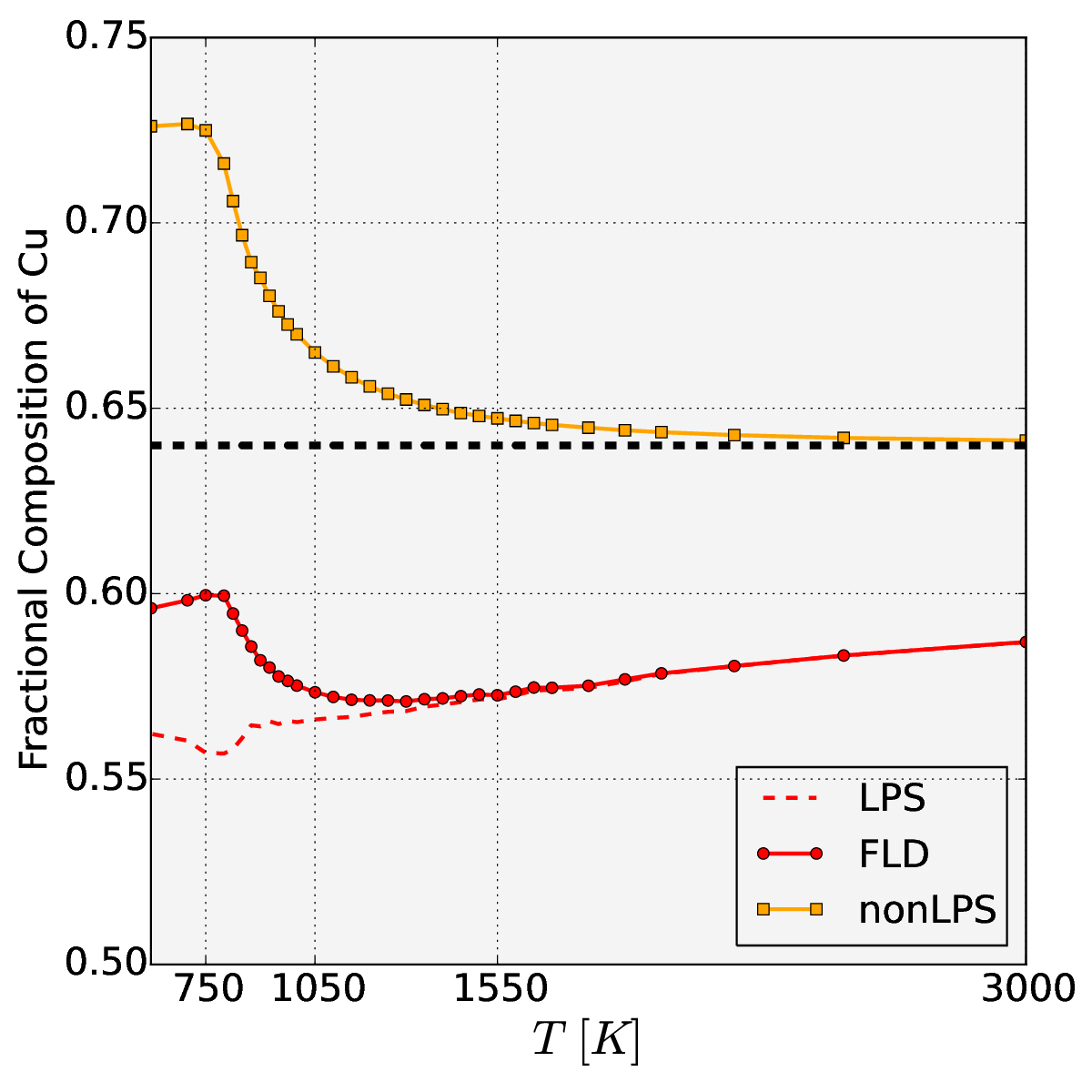}
\caption{The concentration of Cu atoms in: an average 13-atom icosahedron (red dashed-line),
the connected icosahedron domains (red circles),
and in non-icosahedron structures (yellow squares). When approaching $T_{g}$,
the preferential sharing of Zr atoms amongst icosahedra allows
the extensive FLDs to approach the overall alloy composition, while
preserving the Zr-rich environment required by the icosahedra.} \label{fig:comp}
\end{figure}

The size of an FLD is measured by the number of icosahedra
participating in the connected domain. The inset of
Figure~\ref{fig:icos_pop} shows the the weighted average FLD size,
$\bar s$, whose growth accelerates tremendously with decreasing
temperature - note that the data is plotted on a log scale. The
fractional occupation of an average FLD - the proportion of the
icosahedron population in a 30,000 atom system contained in a
typical connected domain - is plotted in Figure~\ref{fig:perc_net}
versus inverse temperature. Here we see that cooling through
$T_{D}$ leads to a dramatic unification of the liqud's FLDs. Above
$T_{D}$ a typical FLD contains a neglible portion of all the
icosahedra, whereas below $T_{D}$ the rapidly growing FLDs begin
join with one another, such that macroscopic number of the
system's icosahedra can be found in a single connected FLD. The
peak value in the susceptibility of the fractional occupation
number, $\frac{d}{dT}(\bar s/N_{icos})$, resides near $900K$ and
is marked $T_{peak}$. By $825K$, $\bar s/N_{icos}$ nearly reaches
unity - only a negligible number of icosahedra can be found to be
separate from a single, extensive network of icosahedra.

These extensive FLDs, we propose, are the underlying structures
responsible for the decoupled dynamics and the enhanced barriers
to diffusion that emerge below $T_{D}$. Indeed, the
disproportionately depressed diffusivity for Zr can be accounted
for by considering the chemical ordering that accompanies the
formation of FLDs. In accordance with efficient packing schemes of
different-sized hard spheres \cite{Sheng2006,Miracle2003},
icosahedra are Zr-rich relative to the overall alloy composition
Cu$_{64}$Zr$_{36}$ - below $T_{A}$ an icosahedron has an
approximate composition of Cu$_{57}$Zr$_{43}$. An earlier MD study
showed that non-interpenetrating connections between the
icosahedra in an FLD strongly favor sharing Zr atoms amongst one
another \cite{Soklaski2013}; for instance, $80\%$ of all shared
vertices are Zr atoms, whereas the icosahedron shell composition
is only $43\%$ Zr. Thus non-interpenetrating connections within
FLDs are utilized to foster a locally Zr-rich environment for the
icosahedra, while the FLDs at large can, necessarily, maintain a
composition closer to that of the alloy. The temperature dependent
Cu concentrations of individual icosahedra (including the center
Cu atoms), FLDs, and the non-icosahedron regions of the liquid are
plotted in Figure~\ref{fig:comp}.

\begin{figure}[t]
\includegraphics[scale=0.38]{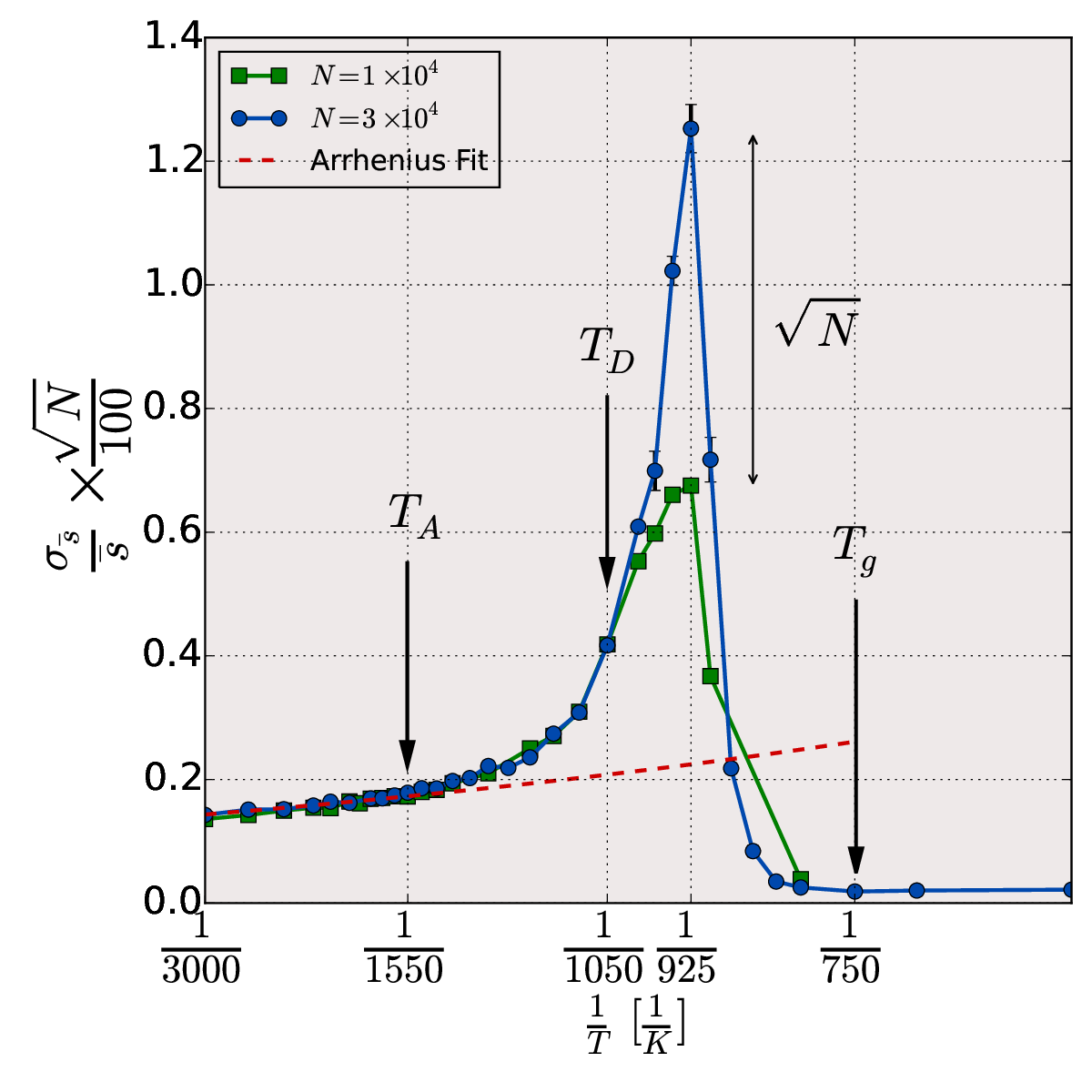}
\caption{Relative fluctuations of the mean connected-icosahedron
domain size over time, as a function of $\frac{1}{T}$, scaled by
$\sqrt{N}$, where $N$ is the number of atoms in the system. A deviation
from the high temperature Arrhenius fit (dashed curve) begins at
$T_{A}$. Domain-size fluctuations grow rapidly upon cooling
through $T_{D}$, and become system size independent. The distinct peak near $925K$
demonstrates the abrupt stabilization and growth of a large FLD,
which percolates above $T_{g}$. Error bars are derived from
averaging two independent runs} \label{fig:netflucs}
\end{figure}

There are numerous studies that demonstrate the unique role played
by icosahedron FLDs in creating strong dynamical heterogeneities
in Cu$_{64}$Zr$_{36}$ as it is supercooled towards $T_{g}$
\cite{Cheng2008,Ding2014,Mendelev2015}. In particular, Y. Zhang
\textit{et al.} recently showed quite explicitly that diffusing
atoms strongly avoid regions of icosahedral ordering, and that
they instead reside in liquidlike regions of the system
\cite{Mendelev2015}. They demonstrate that underlying the development
of medium-ranged icosahedral ordering and glassy dynamics is a crucial correspondence between
dynamical and structural correlation length scales in the liquid.
We thus conclude that the growing barrier to
diffusion and decoupled motions of Cu and Zr atoms, which ensue
below $T_{D}$, is created by the rapidly proliferating, Zr-rich
FLDs of connected icosahedra that form amidst Cu-rich liquidlike
regions in the system.

\section{FLD fluctuations and falling out of equilibrium}

Both cooperative characteristic temperatures, $T_{A}$ and $T_{D}$,
manifest as enhancements in fluctuations of the average domain
size, $\overline{s}$ . Figure~\ref{fig:netflucs} shows the
relative time-fluctuations in domain size,
$\frac{\sigma_{\overline{s}}}{\overline{s}}$, versus inverse
temperature for system sizes of $1\times 10^{4}$ and $3\times
10^{4}$ atoms. The fluctuations are scaled by a factor of
$\sqrt{N}/100$ to make-clear the behavior of system size effects.
Here, $\overline{s}(t)$ is the mean domain size of the system at a
time t, and $\sigma_{\overline{s}}$ is the standard deviation of
$\overline{s}(t)$ from its time-averaged value, $\overline{s}$. As
the liquid is cooled through $T_{A}$, fluctuations in the domain
size become enhanced, growing faster than does the
high-temperature Arrhenius fit (dashed curve) with decreasing
temperature. Above $T_{D}$, the scaled fluctuations for the two
system sizes collapse onto a single curve, indicating that FLD
fluctuations scale as $\sim1/\sqrt{N}$ in this temperature range.
Cooling further through $T_{D}$ leads to a breakdown in this
scaling law, with the peak fluctuations growing as $\sim\sqrt{N}$
on the scaled axis - that is, the peak-value fluctuations appear
to become system size independent. By $T=925K$ the
cooperatively-rearranging FLDs attempt to grow far beyond their
average sizes before inevitably collapsing, reaching relative
fluctuations of 70\%.

This behavior changes abruptly as the liquid is supercooled
through $925K$. Here the FLDs begin to stabilize and the
fluctuations in FLD size drop rapidly with decreasing temperature.
We must consider whether this drop in fluctuations is actually a
mark of structural stabilization or if it is a mere consequence of
our simulation timescale becoming insufficient for capturing
metasable equilibrium dynamics. It is certainly true that the
fluctuations must inevitably fall to zero as we approach $T_{g}$
given that $\tau_{M}$ will far outgrow our simulation timescale.
That being said, $\tau_{M}$ is still sufficiently short at $900K$,
where the fluctuations have already started to fall, that our
liquid has been allowed to evolve for a time four orders of
magnitude longer than the liquid relaxation time, and our data
collection window is two orders of magnitude longer. Furthermore,
several studies show that utilizing slower quench rates, longer
relaxation time, and thermalization techniques only further
enhances the degree of icosahedral ordering and the connectivity
of the resulting FLDs \cite{Ding2014,Mendelev2015}. This indicates
that the drop in fluctuations indeed reflects pronounced
stabilization of the FLDs, and that the fluctuations in the
extensive FLD's size would approach zero even if we were able to
remain in equilibrium near $T_{g}$. By $900K$, nearly $40\%$ of
the system's icosahedra; as we saw in Figure~\ref{fig:perc_net},
this fraction continues to grow rapidly until a single FLD
percolates the system by $825K$ \cite{Soklaski2013},
where the fluctuations nearly
reach zero. A structure is said to percolate a
material when a singly-connected domain of that structure spans the material
(e.g. a connected domain of icosahedra is as long as the diagonal
length of our simulation box). Below $T_{g}$, the percolated FLD shapes the
properties, mechanical and others, of the glass. This has been
demonstrated by a number of excellent studies for several glasses
\cite{Baumer2013, Wu2013, Lee2011, Zhang2014, Wakeda2010, Liu2014,
Wang2014}. Not discussed in this paper are the roles played by the various
types of connections that can form between icosahedra. For detailed discussions
of this important aspect of icosahedral ordering, we refer the reader to early studies
of this subject \cite{Soklaski2013, Ding2014}.

In the limit of large system size, the peaked fluctuations in
Figure~\ref{fig:netflucs} resembles closely the susceptibility
of the fractional occupation plotted in Figure~\ref{fig:perc_net}.
We note that this and other features of the rapid development of extensive
icosahedral ordering in this system indicates the possible
presence of a liquid-liquid phase transition above $T_{g}$
\cite{Tanaka2000}. This will be investigated in detail in a future
publication.

\begin{figure}[t]
\includegraphics[scale=0.6]{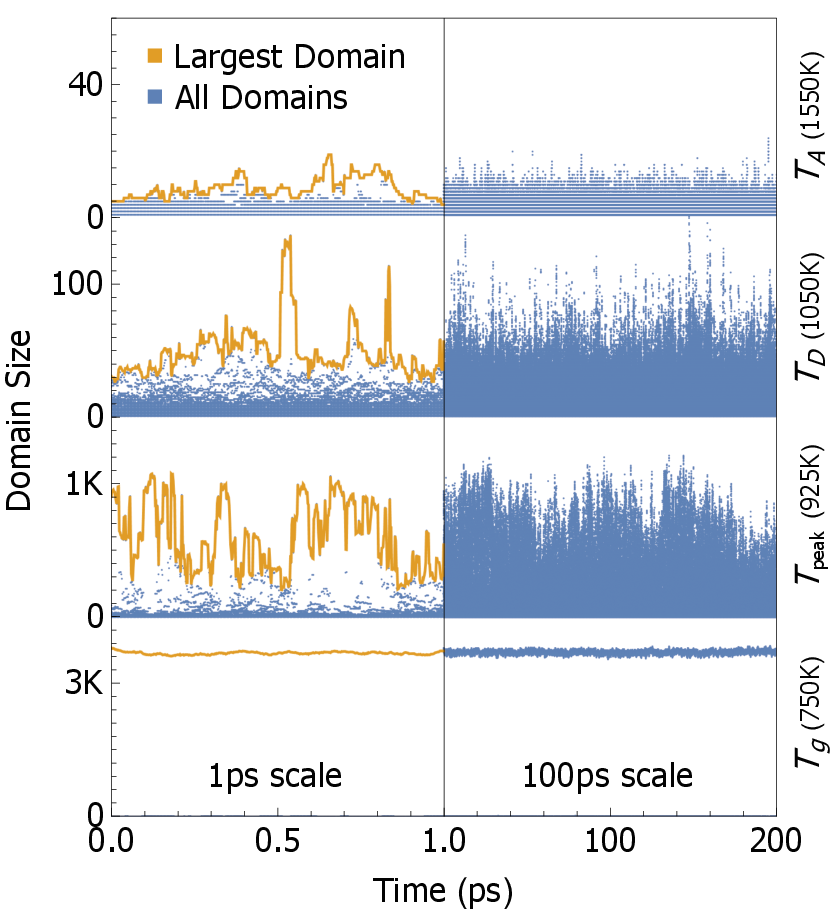}
\caption{(Color Online) The connected-icosahedron domain sizes
present at a given simulation time step plotted versus time.
System temperatures are indicated on the right side of the plots.
On the $1ps$ timescale, an orange trajectory traces the evolution
of the largest domain size.} \label{fig:timeflucs}
\end{figure}

Lastly, the time-dependent behavior of the FLDs that produce the
fluctuations reported in Figure~\ref{fig:netflucs} are shown in
Figure~\ref{fig:timeflucs}. Here, we plot all of the domain sizes
that are present at each simulation time step, for temperatures
$T_{A}$, $T_{D}$, $T_{peak}$, and $T_{g}$. The left side and right
sides of the plot show this time dependence on $1ps$ and $100ps$
timescales, respectively. In the left panel, the evolution of the
largest FLD size is highlighted by an orange trajectory. With
decreasing temperature, down to $T_{peak}$, a clear enhancement in
the time variation of FLD sizes is observed. At $T_{g}$, a single,
large FLD dominates the glass and exhibits minimal fluctuations
across the $200ps$ time window.

\section{Summary}

In summary, we have employed MD simulations to show that distinct
structural features accompany dynamical crossovers in liquid
$\mathrm{Cu_{64}Zr_{36}}$. Comparing the liquid relaxation time,
$\tau_{M}$, to measures of structure lifetimes, $\tau_{LC}$ and
$\tau^{(FLD)}_{icos}$, allows us to identify the temperature
ranges in which cooperative structural rearrangements contribute
to the liquid's relaxation processes. At $T_{A} \approx 2\times
T_{g}$  atoms begin to cooperatively form connected icosahedra off
of isolated ones, serving as the beginning of the icosahedral
ordering that eventually dominates the system. This accompanies
the onset of the super Arrhenius growth of $\eta$ and the break
down of the Stokes-Einstein relationship in the liquid.

In the supercooled regime, $T_{D}\approx 1.4 \times T_{g}$ marks
the early development of stretched glassy dynamics, and the
decoupling of Cu and Zr diffusivities. These dynamical features are a
manifestation of rapidly growing Zr-rich domains of connected
icosahedra, which possess macroscopics numbers of the liquid's
icosahedra. Below $925K$ these FLDs stabilize abruptly, and
eventually percolate the system before the liquid reaches $T_{g}$.
Our results provide needed evidence and explanation for the
structural roles played at $T_{A}$, which according to recent
experimental results is an important characteristic temperature
for all metallic liquids \cite{Kelton2014}, and more broadly
illustrates that a cascade of cooperative structural and dynamical
effects begins at this temperature and characterizes the
supercooled liquid as it approaches $T_{g}$.

\section*{ACKNOWLEDGMENTS}

Fruitful discussions with Takeshi Egami are gratefully
acknowledged. RS, VT, and LY are supported by the National Science
Foundation (NSF) under Grant No. DMR-1207141. KFK is supported by
NSF under Grant No. DMR-12-06707. ZN is supported by NSF under
Grants No. DMR-1106293 and DMR-141122. ZN is grateful to the
Feinberg foundation for visiting faculty program at the Weizmann
Institute. The computational resources have been provided by the
Lonestar and Stampede of Teragrid at the Texas Advanced Computing
Center (TACC) and the Edison cluster of the National Energy
Research Scientific Computing Center (NERSC). All plots were made
using the matplotlib Python library \cite{matplotlib}.

\end{document}